\documentclass{aa} 
\pdfoutput=1
\usepackage{graphicx}
\usepackage{txfonts}
\usepackage{soul}
\usepackage{natbib}

\bibpunct{(}{)}{;}{a}{}{,}
\begin{document} 

   \title{Dissecting stellar chemical abundance space with t-SNE}

   \author{F. Anders\inst{1, 2}, C. Chiappini\inst{1, 2}, B. X. Santiago\inst{3, 2}, G. Matijevi\v{c}\inst{1}, A. B. Queiroz\inst{3, 2}, M. Steinmetz\inst{1}, G. Guiglion\inst{1}
   }
   
   \authorrunning{F. Anders et al.}      
   \titlerunning{t-SNE analysis of the chemistry space of the solar neighbourhood}      
   
     \institute{Leibniz-Institut f\"ur Astrophysik Potsdam (AIP), An der Sternwarte 16, 14482 Potsdam, Germany\\
              \email{fanders@aip.de}
     \and{Laborat\'orio Interinstitucional de e-Astronomia, - LIneA, Rua Gal. Jos\'e Cristino 77, Rio de Janeiro, RJ - 20921-400, Brazil}
     \and{Instituto de F\'\i sica, Universidade Federal do Rio Grande do Sul, Caixa 
Postal 15051, Porto Alegre, RS - 91501-970, Brazil}
	}

   \date{Received \today; accepted ...}

  \abstract
   {
   In the era of industrial Galactic astronomy and multi-object spectroscopic stellar surveys, the sample sizes and the number of available stellar chemical abundances have reached dimensions in which it has become difficult to process all the available information in an effective manner. In this paper we demonstrate the use of a dimensionality-reduction technique (t-distributed stochastic neighbour embedding; t-SNE) for analysing the stellar abundance-space distribution. While the non-parametric non-linear behaviour of this technique makes it difficult to estimate the significance of found abundance-space substructure, we show that our results depend little on parameter choices and are robust to abundance errors. By reanalysing the high-resolution high-signal-to-noise solar-neighbourhood HARPS-GTO sample with t-SNE, we find clearer chemical separations of the high- and low-[$\alpha$/Fe] disc sequences, hints for multiple populations in the high-[$\alpha$/Fe] population, and indications that the chemical evolution of the high-[$\alpha$/Fe] metal-rich stars is connected with the super-metal-rich stars. We also identify a number of chemically peculiar stars, among them a high-confidence s-process-enhanced abundance-ratio pair (HD91345/HD126681) with very similar ages and $v_X$ and $v_Y$ velocities, which we suggest to have a common birth origin, possibly a dwarf galaxy. Our results demonstrate the potential of abundance-space t-SNE and similar methods for chemical-tagging studies with large spectroscopic surveys.}

   \maketitle


\section{Introduction}

One of the major goals of modern Galactic astrophysics is to infer the formation history of our Milky Way. To achieve this aim it is necessary to obtain precise 6D stellar phase-space positions, detailed chemical abundance patterns, and reliable age estimates for large stellar samples. This chrono-chemo-kinematical map of the Galactic stellar populations can then be compared to predictions of various Milky-Way models, eventually unveiling the star-formation and dynamical history of our Galaxy. 

Massive spectroscopic observing campaigns such as RAVE \citep{Steinmetz2006}, SEGUE \citep{Yanny2009}, the Gaia-ESO survey \citep{Gilmore2012}, LAMOST \citep{Deng2012}, APOGEE \citep{Majewski2017}, and GALAH \citep{Martell2017} have in the past decade increased both the volume coverage and the statistical sample sizes by more than two orders of magnitude, to $5\cdot10^6$ stars distributed from the solar vicinity to the far side of the Galactic bulge and the outer halo. 
In spite of this recent conquest of the Milky Way in terms of number of spectroscopically analysed stars, detailed multi-abundance chemo-kinematical studies of the immediate solar vicinity \citep[e.g.][]{Edvardsson1993, Fuhrmann1998, Fuhrmann2011, Fuhrmann2017, Adibekyan2012, Bensby2014, Nissen2015, Nissen2016, DelgadoMena2017} remain at least equally important for Galactic Archaeology (see \citealt{Lindegren2013} for a quantitative analysis). Also, before {\it Gaia} DR2, reliable stellar ages are still mostly confined to the solar vicinity (for exceptions using asteroseismology see \citealt{Chiappini2015a, Martig2015, Casagrande2016, Anders2017, Rodrigues2017, Miglio2017}).

The wealth of new data, especially the high dimensionality of chemo-kinematics space, requires new statistical analysis methods to efficiently constrain detailed Milky-Way formation models (including e.g. stellar evolution, stellar chemical feedback, chemical evolution, and dynamical evolution). Traditionally, the metallicity distribution function and 2D chemical-abundance diagrams ([X/Fe] vs. [Fe/H]), and abundance gradients have been used to constrain the chemical evolution of stellar populations \citep[e.g.][]{Pagel2009}. On the other hand, it is also possible to {\it define} stellar populations by chemistry (e.g. carbon-enhanced metal-poor stars - \citealt{Beers2005}; the chemical thick disc - \citealt{Fuhrmann1998}; high-[$\alpha$/Fe] metal-rich stars - \citealt{Adibekyan2011}), and to then study their structural and chemo-kinematic properties in detail. Abundance-space populations are usually defined in a simple fashion, by dissecting only one 2D abundance diagram. 

More thorough multi-dimensional abundance-space studies using data-mining techniques have emerged over the past years. In a pioneer study, \citet{Ting2012} used principal-component analysis (PCA) to determine the effective dimensionality of abundance space accessible by spectroscopic surveys. \citet{daSilva2012, daSilva2015}, and \citet{Jofre2017} used tree clustering to find groups of stars with similar abundance patterns. Recently, \citet{Boesso2018} studied a solar-vicinity literature compilation and combined hierarchical clustering and PCA to find peculiar chemical subgroups that do not follow the chemical-enrichment flow of the Galactic disc. Their results also suggest that 90\% of the variance in the abundance data can be explained by two principal components that capture the main contributions to chemical enrichment. This is slightly at odds with the earlier work of \citet{Ting2012} who suggest that spectroscopic abundance space has at least an effective dimension of 4.

In this paper we explore the possibility of combining the information contained in various measured abundance ratios using the dimensionality reduction technique t-SNE (t-distributed stochastic neighbour embedding) to define more robust subpopulations and better identify outliers. 
In astronomical applications, t-SNE has mainly been used to identify objects with peculiar spectra (e.g. \citealt{Matijevivc2017, Valentini2017, Traven2017, Anders2018, Reis2018}). \citet{Jofre2017c} employed t-SNE to identify spectral twins in the RAVE database. Recently, \citet{Kos2018} demonstrated in a complementary analysis that t-SNE can also be used as a chemical-tagging tool in chemical-abundance space: the authors were able to recover 7 out of 9 known open and globular clusters with high efficiency and low contamination using 13 chemical abundances from the GALAH survey \citep{Martell2017}, and they also found two new field member stars to known clusters with this technique. 

Here we apply abundance-space t-SNE to the high-resolution solar-vicinity HARPS-GTO survey data of \citet{DelgadoMena2017}, and demonstrate that this method provides a powerful visualisation and clustering tool also for field-star chemical-tagging studies. We identify, in a robust way, several distinct chemical-abundance substructures of the solar-vicinity disc population, as well as some peculiar stars. In a subsequent paper that extends our analysis to other surveys (Chiappini et al., in prep.) we discuss the main result: the detection of distinct chemical sub-populations in the high-[$\alpha$/Fe] regime that points to a different origin of the metal-poor and metal-rich part of the high-[$\alpha$/Fe] disc. 

The paper is structured as follows: Sec. \ref{method} introduces t-SNE. Section \ref{harps} describes the t-SNE results for the high-resolution spectroscopic solar-vicinity survey of \citet{DelgadoMena2017}, considering possible caveats in our analysis and characterising each of the found subpopulations. We finish with a discussion and conclusions in Sec. \ref{conclusions}.

\section{Dissecting chemistry space with t-SNE}\label{method}

Interpreting multi-dimensional abundance distributions determined by spectroscopic surveys is not a trivial task, since different elements originate in different astrophysical sites and at different rates, and their abundance determination is affected by variable observational errors. A convenient way to simplify this problem is dimensionality reduction, i.e. the projection of the N-dimensional abundance space onto a lower-dimensional space in which the chemical similarity between two stars is reflected by their distance in that space. Possibly the best-known such method is PCA, widely used also in astronomical literature. For highly-correlated datasets such as spectral pixel spaces or chemical-abundance spaces, however, more sophisticated non-linear methods like IsoMap or locally linear embedding are known to perform much better \citep[e.g.][]{Matijevivc2012, Ivezic2013}.

In this paper, we reanalyse the high-resolution spectroscopic solar-vicinity survey of \citet{DelgadoMena2017} using a machine-learning algorithm called t-distributed stochastic neighbour embedding \citep[t-SNE;][]{Hinton2003, vanderMaaten2008}. This method is widely used in big-data analytics, and is able to efficiently project complex datasets onto a 2D plane in which the proximity between similar data points is preserved. We use the python implementation of t-SNE included in the {\tt scikit-learn} package \citep{Pedregosa2012} and refer to the original papers and the online documentation for details about the method and code. In short, the advantage of using t-SNE over other manifold-learning techniques is that it performs much better in revealing structure at many different scales \citep{vanderMaaten2008, Matijevivc2017}, which is a necessary feature when looking for chemical substructure in the Galactic disc.

\begin{figure*}\centering
 \includegraphics[width=0.96\textwidth]{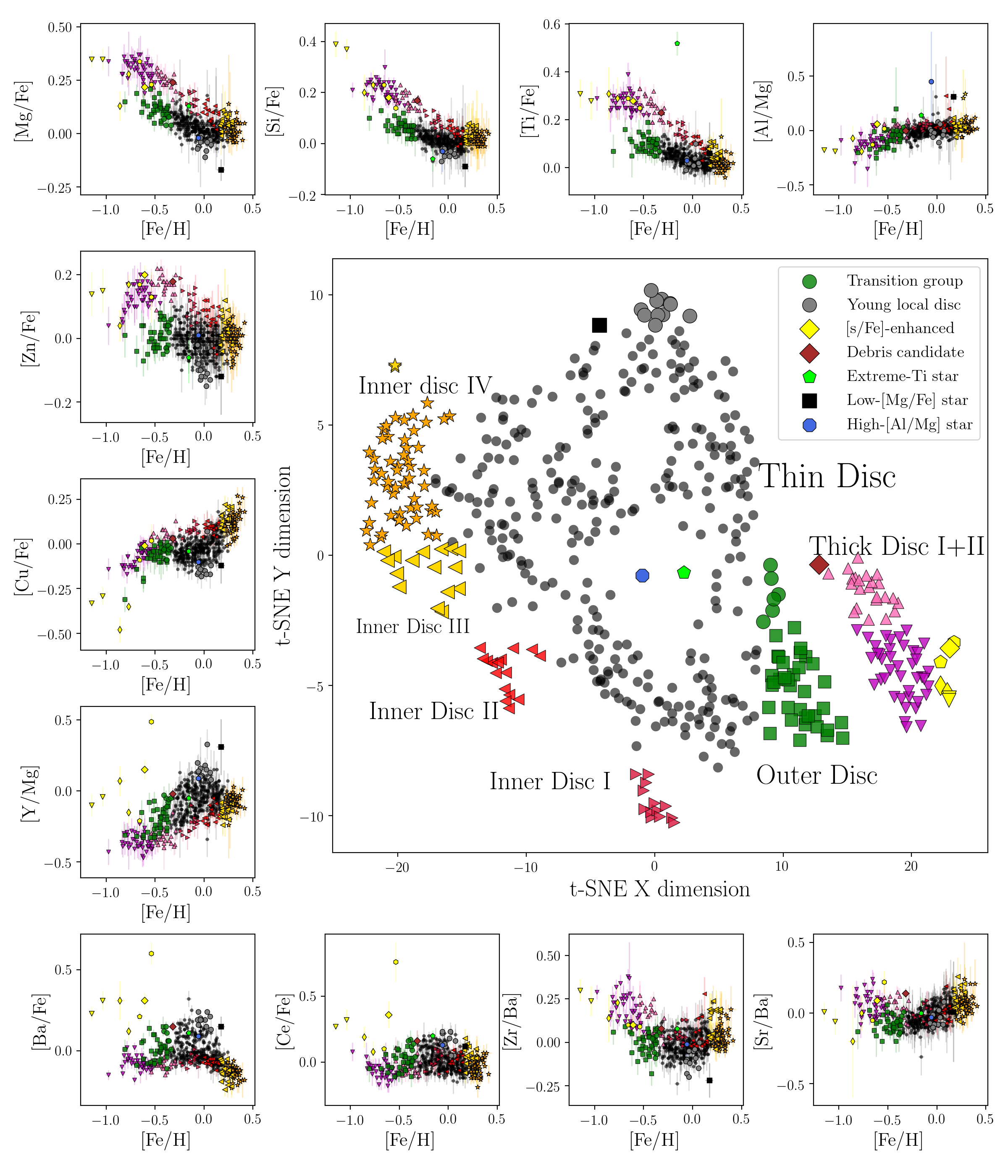}
\caption{Illustration of how t-SNE works in abundance space, using the \citet{DelgadoMena2017} sample. The small panels show eleven of the possible $\sim20,000$ abundance diagrams that can be created from 13 elements. The resulting reference t-SNE projection of the full abundance space is shown in the big panel, and several identified subgroups are indicated. For reference, the Sun would be found very close to the solar-abundance twin HD212036 ([Fe/H]=-0.01, [X/Fe]=0+-0.05 for all measured elements X) that is located at ($X_{\rm t-SNE},Y_{\rm t-SNE}) = (-4.57,1.46)$.}
\label{harps0}
\end{figure*}

{\it How t-SNE works:} For a given set of $N$ high-dimensional datapoints $\mathbf{x}_1, \dots, \mathbf{x}_N$ (images, spectra, or in our case chemical-abundance vectors), t-SNE first computes pairwise similarity probabilities $p_{ij}$ for the points $\mathbf{x}_i$ and $\mathbf{x}_j$:
$$p_{j\mid i} = \frac{\exp(-\lVert\mathbf{x}_i - \mathbf{x}_j\rVert^2 / 2\sigma_i^2)}{\sum_{k \neq i} \exp(-\lVert\mathbf{x}_i - \mathbf{x}_k\rVert^2 / 2\sigma_i^2)}.$$
To circumvent problems with outliers, the symmetrised similarity of $x_j$ and $x_i$ is defined as 
$$p_{ij} = \frac{p_{j\mid i} + p_{i\mid j}}{2N}.$$

In the next step, t-SNE attepts to learn a $d$-dimensional map $\mathbf{y}_1, \dots, \mathbf{y}_N$ (in general $d=2$) that reflects the similarities  $p_{ij}$ similarities between two points $\mathbf{y}_i$ and $\mathbf{y}_j$ in the low-dimensional map, defined as
$$q_{ij} = \frac{(1 + \lVert \mathbf{y}_i - \mathbf{y}_j\rVert^2)^{-1}}{\sum_{k \neq m} (1 + \lVert \mathbf{y}_k - \mathbf{y}_m\rVert^2)^{-1}}.$$
This metric uses Student's $t$ distribution to avoid crowding problems in the low-dimensional map \citep{vanderMaaten2008}. Starting from a random Gaussian distribution in the $d$-dimensional map, the locations of the points $\mathbf{y}_i$ are determined by minimizing the Kullback–Leibler divergence \citep{Kullback1951} between the low- and high-dimensional similarity distributions $Q$ and $P$:
$$KL(P||Q) = \sum_{i \neq j} p_{ij} \log \frac{p_{ij}}{q_{ij}},$$
using a gradient-descent method. The result of this optimization is a 2D (or 3D) map that reflects the similarities between the high-dimensional inputs (see Figs. \ref{harps0} and \ref{harps1}).

The method has one main parameter, the so-called perplexity, $p$, which governs the bandwidth of the Gaussian kernels $\sigma_i$ appearing in the similarities $p_{ij}$. As a result, the bandwidth is adapted to the density of the data: smaller values of $\sigma_i$ are used in denser parts of the data space. The perplexity parameter can be thought of as a guess about the number of close neighbors each point has, and therefore the ideal value for $p$ depends on the sample size. A change in perplexity has in many cases a complex effect on the resulting map, and different values for $p$ should be explored \citep{Wattenberg2016}. 

Recently, \citet{Linderman2017} demonstrated that two other hyper-parameters of t-SNE can be chosen optimally: the learning rate should be set to $\sim 1$, and the early-exaggeration parameter should be set to $\sim 0.1$ times the sample size. In the following, we use these recommendations.

In addition, t-SNE, as a genuine machine-learning technique, does have two drawbacks that are relevant for our science case. First, it does not account for individual uncertainties, and may therefore be affected by extremely heteroscedastic errors. We mitigate this shortcoming by performing a simple Monte-Carlo experiment (Sec. \ref{robust}) to show that our results are robust to abundance uncertainties. Secondly, its current implementations do not allow to treat missing data, so that any star with a missing individual abundance measurement has to be excluded. We therefore decided to focus on the most inclusive set of chemical abundances (see Sec. \ref{harps}).

\section{Re-analysing the HARPS GTO sample}\label{harps}

\begin{figure*}\centering
 \includegraphics[width=0.99\textwidth]{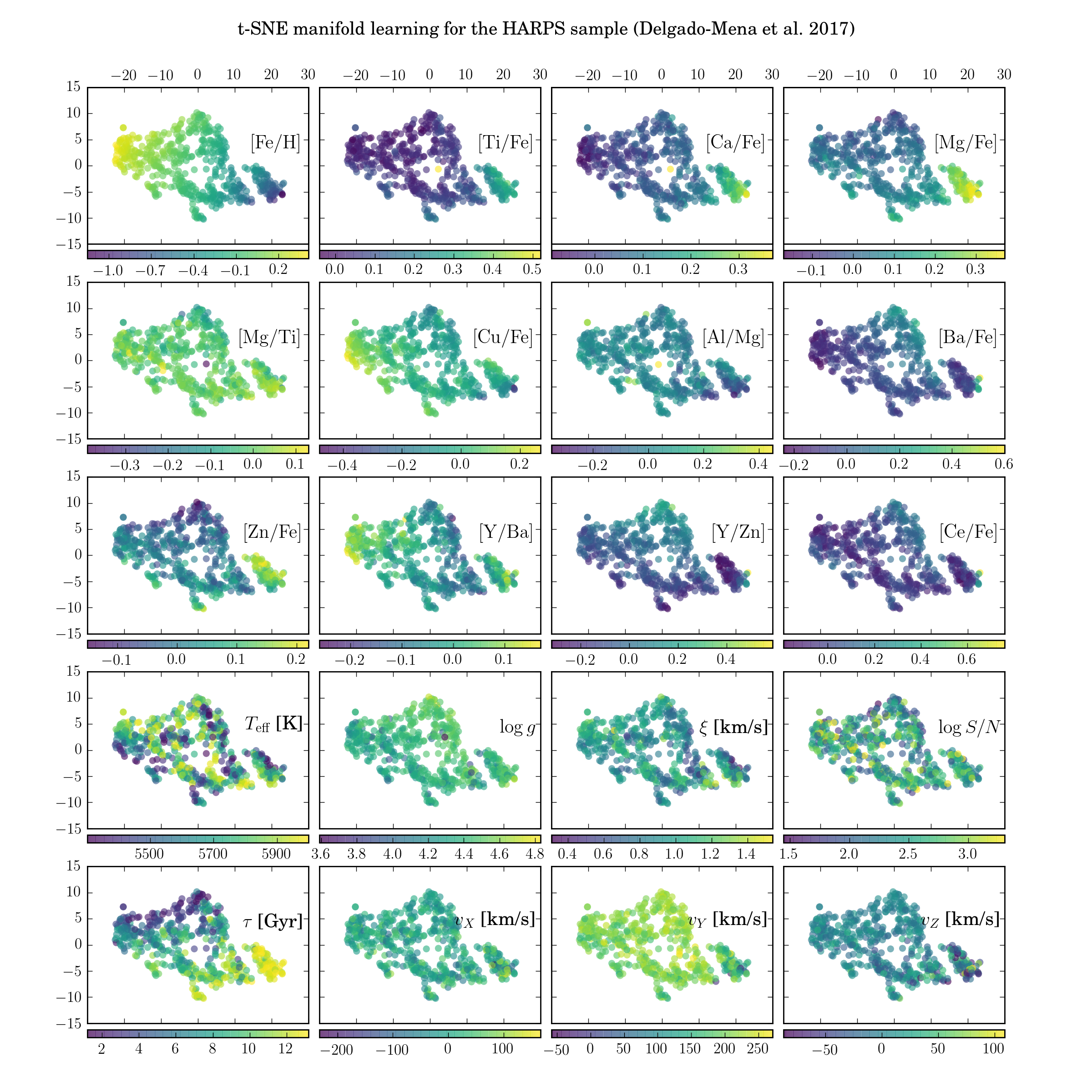}
\caption{Fiducial t-SNE projection of the \citet{DelgadoMena2017} sample (see big panel in Fig. \ref{harps0}), colour-coded by chemical abundances (top three rows), spectroscopic parameters (effective temperature $T_{\rm eff}$, surface gravity $\log g$, microturbulence $\xi$, and signal-to-noise ratio $\log S/N$; fourth row), age $\tau$ (fifth row, first panel) and space velocities (fifth row). We note that only [Fe/H] and the [X/Fe] ratios were used as input for the t-SNE run.}
\label{harps1}
\end{figure*}

In an extensive series of papers, \citet{Adibekyan2011, Adibekyan2012, DelgadoMena2014, DelgadoMena2015, BertrandeLis2015, Suarez-Andres2017, DelgadoMena2017, DelgadoMena2018} studied the chemical abundances of a sample of 1111 solar-vicinity FGK stars using the very high resolution of the HARPS spectrograph ($R\sim 115,000$). This sample mostly contains metal-rich warm dwarf and subgiant stars, but also includes a wide range of effective temperatures, gravities and metallicities. The HARPS sample initially served to detect and characterise exoplanets is volume-complete within 60 pc and was selected in such a way that metallicity biases are avoided \citep{Adibekyan2013}. The HARPS metallicity distribution (MDF) agrees well with the MDF of the 25pc volume-complete sample of \citet{Fuhrmann2011} and the high-quality local ($d<1$ kpc) APOGEE DR10 red-giant sample of \citet{Anders2014}.

\citet{DelgadoMena2017} recently reanalysed this sample, employing a revised linelist \citep{Tsantaki2013}, improving the effective temperature calibration, and correcting spectroscopic gravities using the {\it Hipparcos} parallaxes of \citet{vanLeeuwen2007}. They report chemical abundances for Mg, Al, Si, Ca, Ti, Fe, Cu, Zn, Sr, Y, Zr and Ba for 1059 stars (Ce, Nd and Eu are available for a substantial subset of these), derived using standard Local Thermodynamic Equilibrium (LTE) analysis using ARES \citep{Sousa2007, Sousa2015} to measure equivalent widths and MOOG \citep{Sneden1973} to measure abundances by comparing to Kurucz ATLAS9 atmospheres \citep{Kurucz1993}. These chemical abundances were complemented by photometry from APASS DR9 \citep{Henden2014} and 2MASS \citep{Cutri2003}, and by astrometry (parallaxes, proper motions) from the {\it Gaia} DR1/TGAS catalogue \citep{Michalik2015, GaiaCollaboration2016}, or when these were unavailable (135/1059 stars), from the re-reduced {\it Hipparcos} data \citep{vanLeeuwen2007}. Using the combined spectroscopic, photometric, and astrometric data, we computed precise stellar masses, ages, distances, and extinctions using the {\tt StarHorse} code \citep{Queiroz2018}. For this run, we employed a fine grid ($\Delta \log\tau=0.01$ dex, $\Delta$[Z/H]=0.02 dex) of PARSEC 1.2S stellar models \citep{Bressan2012, Tang2014, Chen2015}, which significantly improved the precision of our ages with respect to the default grid ($\Delta \log\tau=0.05$ dex, $\Delta$[Z/H]=0.05 dex). The median age precision of the final t-SNE sample is 14\%. 

The kinematic results are based on {\it Gaia} DR1/TGAS positions and proper motions, radial velocities from \citet{Adibekyan2012}, our {\tt StarHorse} distances, and the orbit-integration tool {\tt galpy} \citep{Bovy2015}, using the new Staeckel approximation implementation of \citet{Mackereth2018} to determine orbital eccentricities $e$ and maximum heights above the plane $Z_{\rm max}$. Galactic space velocities were estimated adopting a solar Galactocentric distance of 8.3 kpc, a local standard-of-rest velocity of 220 km/s, and solar peculiar velocities as in \citet{Piffl2014}.

In this section we test the performance of abundance-space t-SNE on this most recent HARPS GTO sample compilation. The high number of measured abundances, in conjunction with the high precision of the measurements and the easily tractable sample size, makes the HARPS sample an ideal test case for machine-learning algorithms. 
Our first tests showed that, in order to obtain reliable t-SNE abundance maps, the sample needed to be analysed in a more restricted temperature range, because certain abundance trends seem to be dominated by underlying temperature trends. Therefore, similar to \citet{DelgadoMena2017}, we chose an effective temperature range of 5300 K $<T_{\rm eff}<$ 6000 K (satisfied for 539 stars) for our analysis. We furthermore excluded one star with $\log g_{\rm HIP}<3$, and required successful abundance determination for Mg, Al, Si, Ca, TiI, Fe, Cu, Zn, Sr, Y, ZrII, Ce and Ba that we use as input for t-SNE, leaving us with 533 stars.\footnote{Carbon and oxygen abundances are available from previous studies \citep{Suarez-Andres2017, BertrandeLis2015}, but since they are based on previous stellar parameter estimates, we decided not to include them in the t-SNE runs and only use them in the interpretation. We also did not use Nd and Eu in the t-SNE run, because they were only available for about half of the sample (stars with the highest signal-to-noise ratios).} To compensate the fact that t-SNE does not take into account individual (heteroscedastic) uncertainties in the data, we followed the approach of \citet{Hogg2016} and rescaled each abundance by the median uncertainty in that element, assuming an abundance uncertainty floor of 0.03 dex. In our final sample of 530 stars we also discarded 3 stars for which our age determination code, {\tt StarHorse} \citep{Santiago2016, Queiroz2018}, did not converge. We verified that these choices do not significantly affect the resulting t-SNE maps. 

The result of our reference t-SNE projection (perplexity $p=40$) is illustrated in Fig. \ref{harps0}. The figure shows how the neighbourhood of points in several abundance diagrams (small panels) is reflected in the t-SNE map (big panel). In Fig. \ref{harps0} we also identified and named some  substructures that clearly emerge from the t-SNE projection. The groupings were visually defined by jointly analysing t-SNE projections with different hyper-parameters (most importantly, perplexity). The robustness of each of the substructures and groups iś discussed extensively below. The naming of the subgroups was inspired by previous knowledge about the chemo-chrono-kinematic relations of Galactic stellar populations.

Figure \ref{harps1} again shows our reference t-SNE map for the HARPS sample, but now colour-coded by chemical-abundance ratios, stellar parameters, ages and kinematics. The panels in the first three rows show how t-SNE is grouping the stars with similar abundances in the two-dimensional plane. The panels coloured as a function of stellar parameters demonstrate that the sample is not subject to major systematic abundance shifts, but does show some residual trends with effective temperature, since it preferentially groups cooler stars in slightly different regions of the t-SNE map than hotter ones. Because part of this effect may be due to chemical evolution rather than systematic abundance errors, we refrained from applying ad-hoc corrections to the abundances. 

\begin{figure}
\sidecaption
 \includegraphics[trim=0cm 2cm 0cm 0cm, clip=true, width=0.53\textwidth]{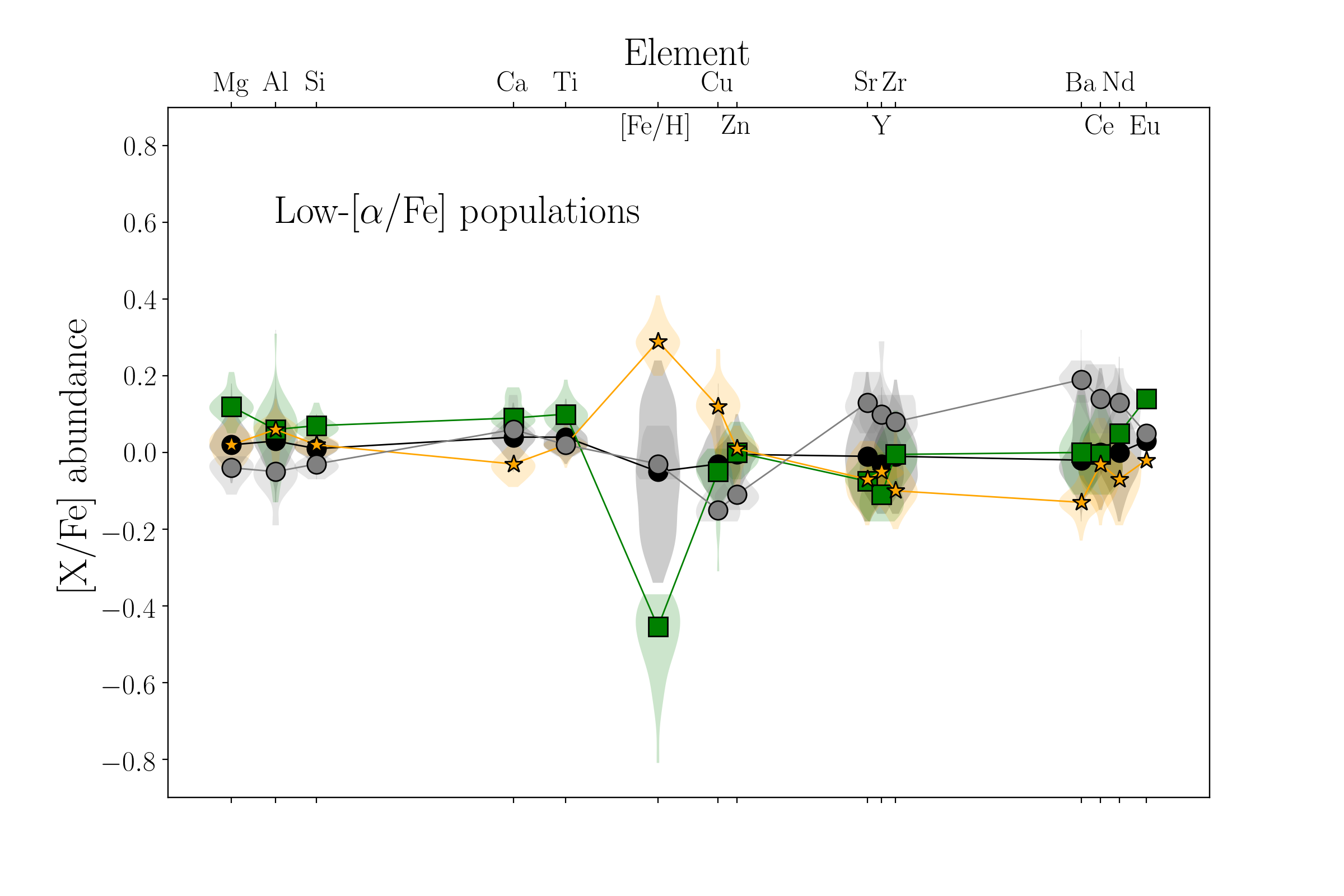}\\
 \includegraphics[trim=0cm 2cm 0cm 2cm, clip=true, width=0.53\textwidth]{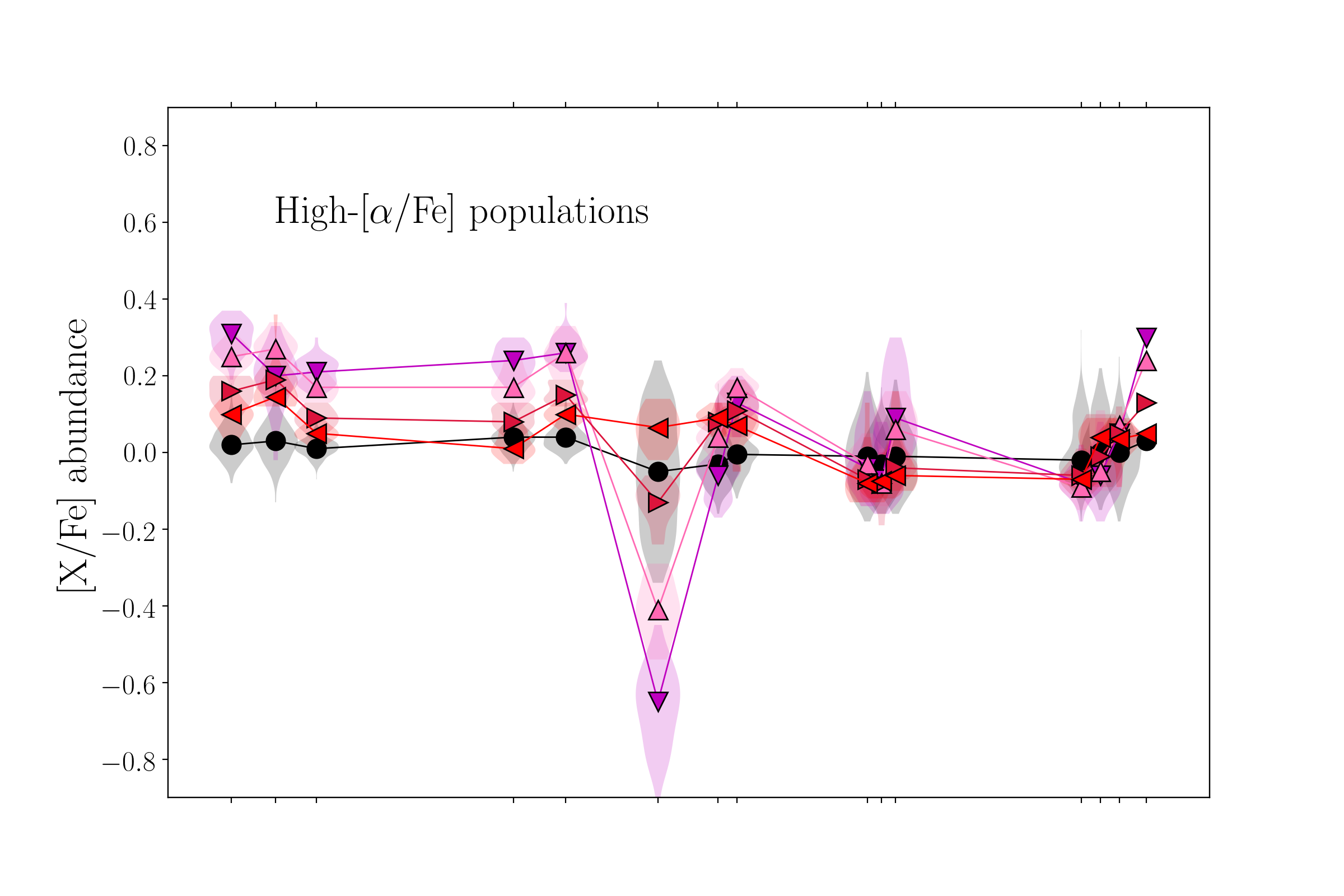}\\
 \includegraphics[trim=0cm 0 0cm 2cm, clip=true, width=0.53\textwidth]{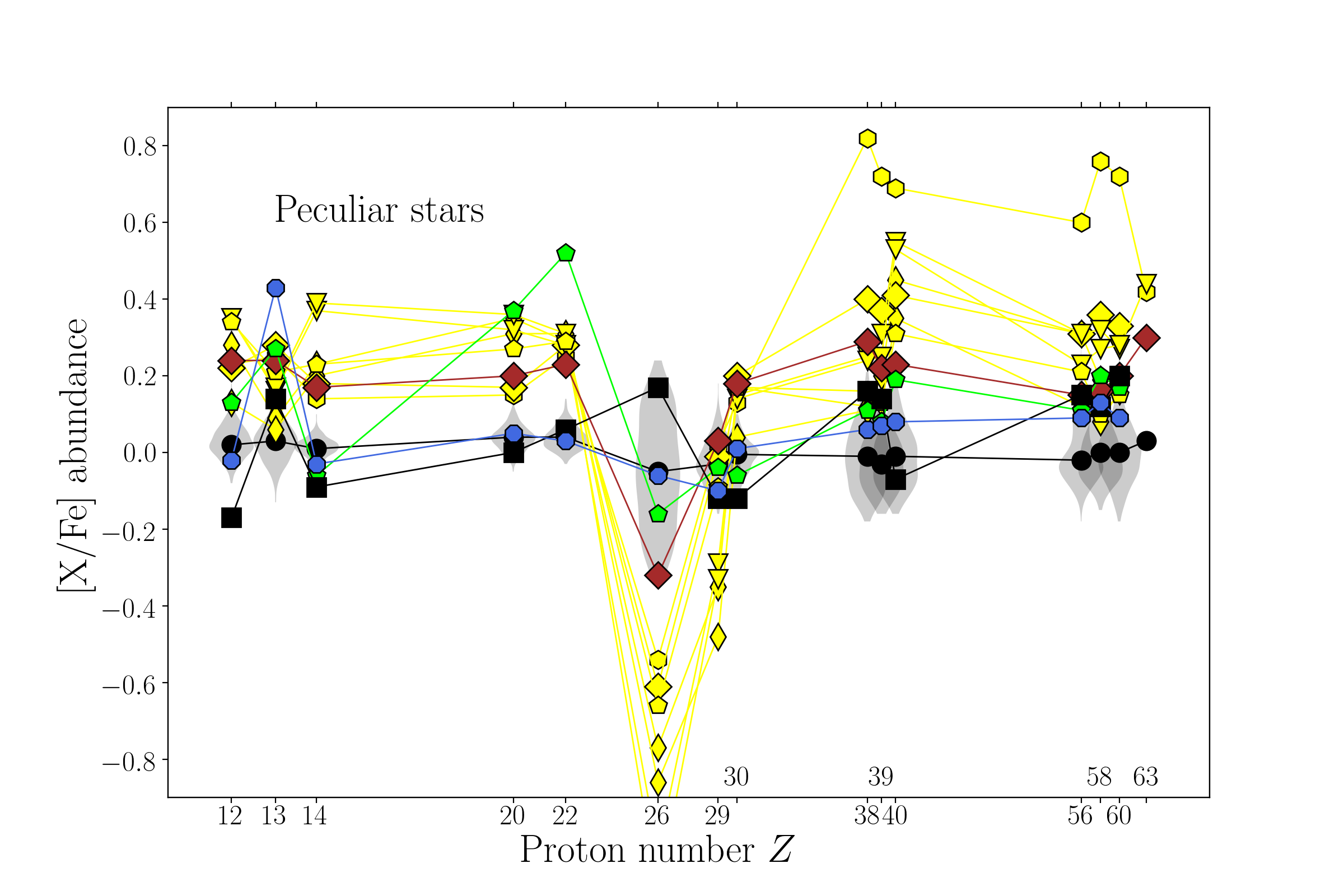}
\caption{Chemical-abundance patterns relative to iron for the t-SNE-selected subsamples of the HARPS survey, using the same symbols and colours as in Fig. \ref{harps0}. For each population we show the median abundance trend, as well as the full abundance distribution (for the top two panels). For visibility, we divided the sample into three groups that are shown separately in the three panels. The ``thin disc'' population (black circles) is shown in all panels for comparison.}
\label{harps3}
\end{figure}

Figure \ref{harps3} shows the corresponding [X/Fe] abundance trends versus proton number for each of the substructures identified in Fig. \ref{harps0}. We now proceed to the discussion of these results.

\subsection{The overall appearance of the t-SNE map}

Our reference t-SNE projection shown in Figs. \ref{harps0} and \ref{harps1} reveals significant amounts of substructure in the local chemical-abundance space. The non-linearity of the method makes it difficult to attribute the overall appearance of the map to specific elemental abundances, which is why we limit this discussion to a qualitative level. In accordance with earlier studies of the dimensionality of abundance space (e.g. \citealt{Ting2012, Boesso2018}), our results suggest that most of the variance of the data is in the metallicity and [$\alpha$/Fe] abundance dimensions, corresponding to the different time-scales of supernovae type Ia and type II. In fact, the X dimension of the t-SNE map correlates very well with metallicity (Pearson's correlation coefficient $r=-0.95$; see Fig. \ref{harps1}, top left panel), which means that a lot of information about the chemical pattern of a star is already given by its metallicity. The t-SNE X dimension is also highly correlated with [$\alpha$/Fe] (e.g., $r=0.92$ for [Ca/Fe] and $r=0.87$ for [TiI/Fe]). 

Figure \ref{harps1} also demonstrates that the t-SNE map's Y dimension, although it also correlates with [$\alpha$/Fe] and [Zn/Fe] abundances, encodes information on s-process abundances, e.g. [Ba/Fe] and [Y/Zn], and consequently stellar age, a variable that was not included in the inference. In principle this opens up the possibility for calibrating multi-element chemical clocks.

The fourth row of Fig. \ref{harps1} also shows that the t-SNE projection responds to elemental-abundance trends with stellar parameters, although they have not been included as input parameters, and although we work in a narrow effective-temperature bin: t-SNE places stars with slightly different stellar parameters in slightly different places of the map. For example, we see some residual abundance trends with $T_{\rm eff}$ (fourth row, left panel), which may either be due to possible systematic abundance errors (see also \citealt{DelgadoMena2017}, or due to real stellar population trends (with stellar mass). In the case of $\log g$ (forth row, second panel), the trends are likely not due to systematic errors, but due to stellar and chemical evolution: at fixed $T_{\rm eff}$ on the main sequence, $\log g$ is a proxy for stellar age, and the abundance patterns are expected to vary with age.

By construction, t-SNE clusters similar-abundance stars in different places of the map. The several discernible islands on the map suggest that we are able to identify stars that were formed from gas with significantly different chemical enrichment than the bulk of the disc stars that live on the ``main island'' of the map. In the following subsection, we will show that most of the substructures identified in Fig. \ref{harps0} are robust to abundance uncertainties and reasonable variations in our analysis.

\subsection{The robustness of the t-SNE results}\label{robust}

\begin{figure}\centering
 \includegraphics[width=0.49\textwidth]{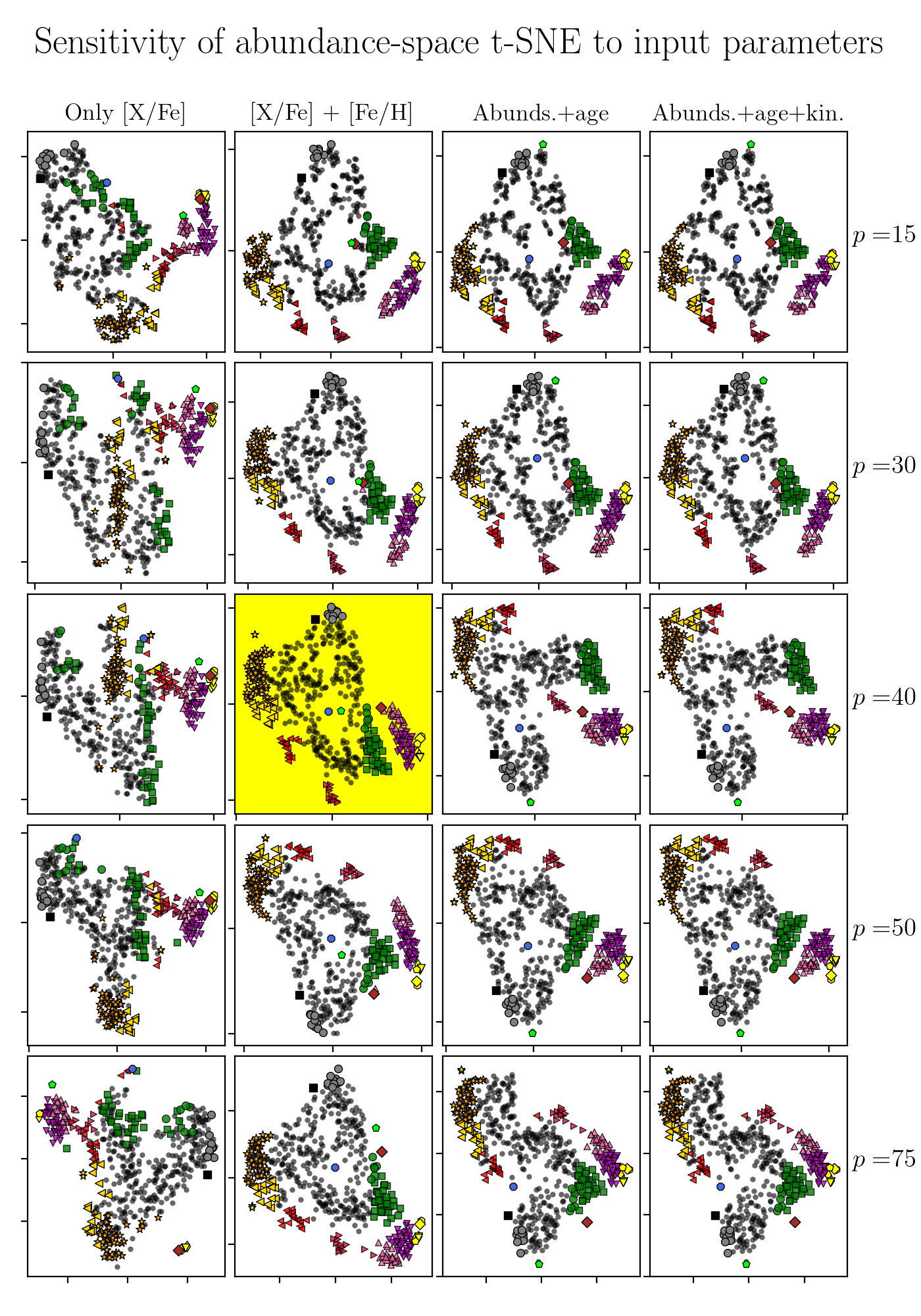}
\caption{t-SNE representations of the chrono-chemo-kinematics space spanned by the \citet{DelgadoMena2017} sample. Each column represents a combination of input information (only [X/Fe], [X/Fe]+[Fe/H], [X/Fe]+[Fe/H]+age, and [X/Fe]+[Fe/H]+age+space velocities, respectively), while each row corresponds to a particular perplexity value, as indicated on the right side of the figure. The panel highlighted in yellow represents the results that we analyse in detail in this paper by defining chemical subpopulations based on this map.}
\label{perplexitytest2}
\end{figure}

As discussed in Sec. \ref{method}, the overall appearance of the maps produced by t-SNE depend mainly on the perplexity parameter $p$, as well as on the chosen parameter space. In Fig. \ref{perplexitytest2}, we show the t-SNE maps for different perplexity values and different sets of input parameters, using the same colours and symbols as in Fig. \ref{harps0}. This experiment shows that: 
\begin{enumerate}
 \item The main features (i.e. neighbourhood relations between points) of the map are preserved (modulo map rotations/reflections) for a wide range of perplexities,.
 \item The groups defined in Fig. \ref{harps0} are also robustly recovered for different perplexities.
 \item Using only [X/Fe] abundance ratios results in slightly different maps, which can be explained by the higher abundance precision of [Fe/H] with respect to the [X/Fe], and the thus higher weight of this dimension in the t-SNE projection. The [Fe/H] dimension alone, however, is not responsible for the emergence of the prominent subgroups.
 \item Adding ages and/or kinematics to the input parameter space does not significantly improve the t-SNE projection, at least in this special case of very local, high-resolution, and high-signal-to-noise data. 
 In the case of moving groups or globular clusters, however, adding kinematic dimensions to chemical tagging exercises does seem to help the recovery of known clusters \citep{Chen2018}.
\end{enumerate}

\begin{figure*}
\sidecaption
 \includegraphics[width=0.64\textwidth]{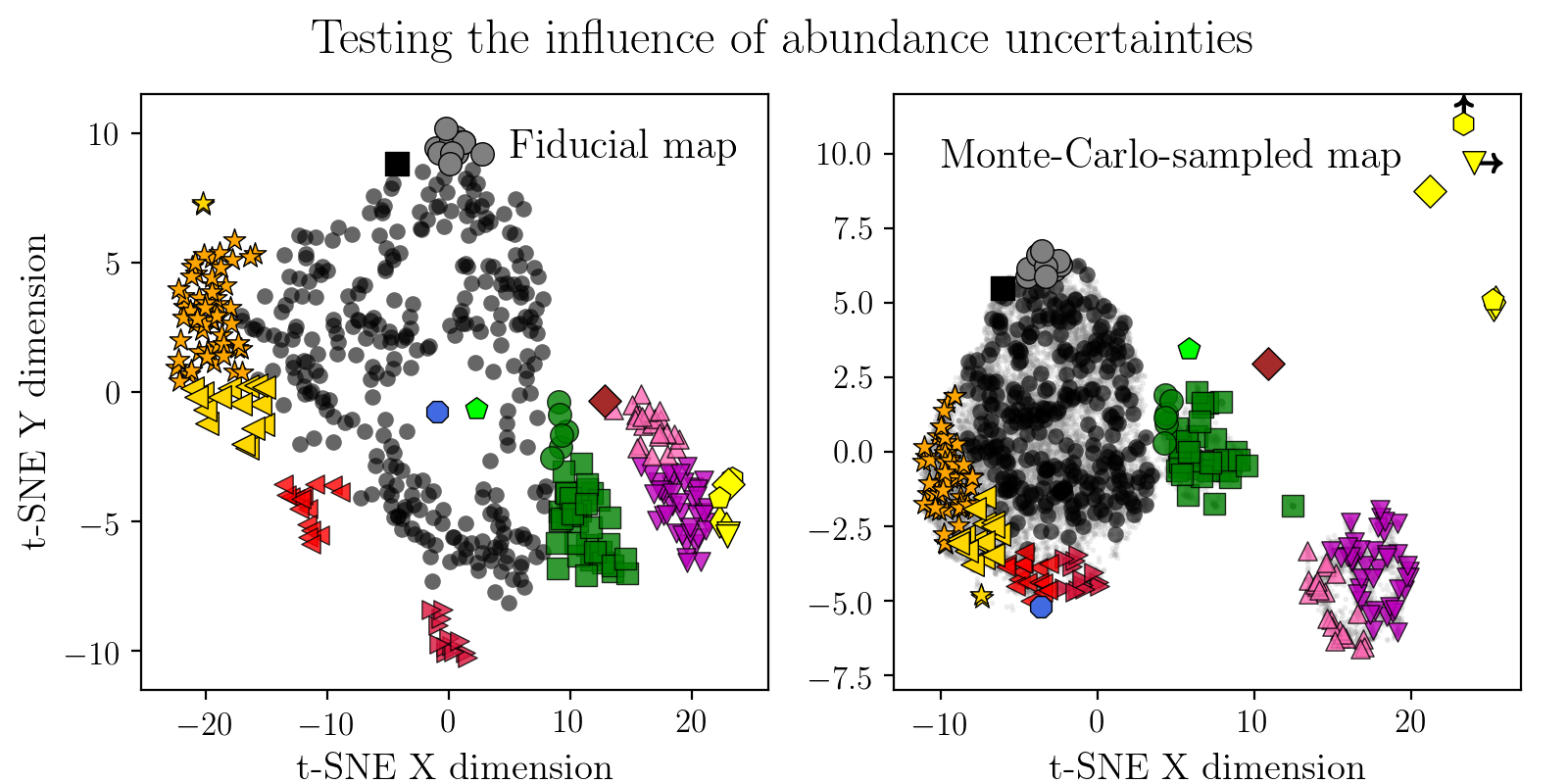}
\caption{Robustness test of our t-SNE-selected subsamples to abundance errors. The left panel shows the fiducial map, while the right panel shows the result of our Monte-Carlo test. For each star, 50 random stars were drawn from a Gaussian centered on the measured abundance, and with dispersions corresponding to the measured uncertainties. t-SNE was then run on this artificially increased sample. The resulting map (for $p=40$) shows the positions of each Monte-Carlo realisation as faint grey dots, and the median position of each star as the same big symbol as in the fiducial map. The experiment demonstrates that most of our selected subgroups are robust to doubling the observational errors of the HARPS sample.}
\label{mctest}
\end{figure*}

We further tested the robustness of our reference map to abundance errors with a simple Monte-Carlo experiment (see Fig. \ref{mctest}): For each star, we created 50 mock stars with abundances drawn from a multi-dimensional Gaussian distribution centered on the measured abundance, and variance corresponding to the measured abundance uncertainties. t-SNE was then run on this artificially increased sample, again with various perplexity value. Because t-SNE cannot take into account uncertainties in the data, this procedure was used to assure that the groups that we identified in the t-SNE map in Fig. \ref{harps0} were not due to chance groupings. This experiment can be regarded as a noise-injection experiment: adding uncertainties to measured (i.e. already noisy) data values blurs the "true" abundance values even more, and makes it even harder to find abundance groups or outliers. This means that if a group/outlier disappears in our Monte-Carlo test, this test does not rule out its existence. On the other hand, if the signal persists, it is very unlikely to be due to a chance grouping. The robustness test shown in Fig. \ref{mctest} was also used for the definition of the chemical populations discussed in the next section.

\subsection{Disc sub-populations}\label{pops}

In this subsection, we discuss the main groups and features identified in Fig. \ref{harps0} in more detail.\\

{\it The thin-thick disc dichotomy:} As discussed in the works of \citet{Adibekyan2011, Adibekyan2012} and \citet{DelgadoMena2017}, the HARPS-GTO data confirm the clear discontinuity between the high- and the low-[$\alpha$/Fe] sequences in the [Mg/Fe] vs. [Fe/H] diagram (e.g. \citealt{Edvardsson1993, Fuhrmann1998, Fuhrmann2011, Fuhrmann2017}). This discontinuity is reflected in a very clear manner in the t-SNE projection:
We find a clear and obvious gap between the chemical thin- and thick-disc populations in the t-SNE diagram that remains very robust for different choices of the t-SNE hyper-parameters. Primarily, this means that the chemical patterns of thin and thick disc are indeed distinct, and can be disentangled by high-resolution spectroscopy. Secondly, our analysis of the full chemical information results in a much more accurate division of the chemically-thin and thick populations. Indeed, if one only relies on one diagnostic, such as the [Mg/Fe] vs. [Fe/H] diagram \citep{Adibekyan2011, DelgadoMena2017}, some thick-disc stars would (probably incorrectly) be identified as belonging to the chemical thin disc (see Fig. \ref{harps0}).

{\it High-[$\alpha$/Fe] sub-populations:} \citet{Adibekyan2011} first discovered a clear discontinuity between the metal-poor and metal-rich $[\alpha$/Fe]-enhanced disc populations (although there are earlier indications in the literature, e.g. \citealt{Fuhrmann2008}, Fig. 30). In our t-SNE analysis of the \citet{DelgadoMena2017} sample, similar to the original paper, we also see a clear difference between at least two, maybe three or four populations (dubbed Thick Disc I/II and Inner Disc I/II in Fig. \ref{harps0}). Even if ages and/or kinematics are included as additional dimensions in the analysis, this picture does not change much. The implications of this result, which we can also confirm with other high-resolution data covering larger volumes, will be discussed in depth in a companion paper (Chiappini et al., in prep.).

The middle panel of Fig. \ref{harps3} shows the abundance profile with respect to iron for each of the four [$\alpha$/Fe]-rich populations, compared to the chemical thin disc. The figure suggests that the most of the abundance variance among the four groups can be captured by one parameter (e.g. metallicity). There are, however, subtle deviations from this pattern: for example, group Thick Disc II is more enhanced in [Al/Fe] than group Thick Disc I. Figure \ref{age} shows that the Inner Disc populations I-IV form a rather tight chemical-evolution sequence in many abundance vs. age diagrams, separated from the Thick-Disc I/II by a jump in [$\alpha$/Fe] abundances.

{\it Super-metal-rich stars:} The Inner Disc IV population (western-most stars in the t-SNE plane; orange stars in Fig. \ref{harps0}) encompasses super-metal-rich stars (SMR; [Fe/H] $\gtrsim0.3$; see \citealt{Grenon1972, Grenon1989, Grenon1999, Chiappini2009}). They have only slightly different abundance patterns from the bulk of the thin-disc stars (black dots; see Fig. \ref{harps3}, top panel); however, Fig. \ref{harps1} shows that they are enhanced in [Y/Ba] and [Cu/Fe] with respect to the local thin disc, indicative of an origin in the inner Milky-Way disc. Figures \ref{age} and \ref{kin} show that most of these stars have ages between 4 and 8 Gyr \citep{Trevisan2011, Casagrande2011, Anders2017}, and are on cold orbits ($e<0.12$; e.g. \citealt{Kordopatis2015}), which again supports the idea that they have radially migrated from the inner disc (see e.g. \citealt{Minchev2012, Minchev2013, Minchev2014, Vera-Ciro2014, Grand2016}).

{\it The transition from h$\alpha$mr to SMR stars:} Most literature measurements agree that the high- and low-[$\alpha$/Fe] sequences in the [$\alpha$/Fe] vs. [Fe/H] diagram merge at super-solar metallicities (e.g. \citealt{Adibekyan2011, Anders2014, Hayden2015}). In other words, the upper metallicity limit of the high-[$\alpha$/Fe]/h$\alpha$mr population is not yet firmly established. Our analysis shows that even when including the full chemistry information, the high-[$\alpha$/Fe]-like and SMR population still form a sequence in the t-SNE projection (e.g. the dark-yellow triangles in Fig. \ref{harps0} have intermediate characteristics between the red triangles and the orange stars). This is why we named the h$\alpha$mr groups ``Inner Disc I/II'' in Fig. \ref{harps0}, meaning that they do not belong to the genuine thick disc. In fact, they are slightly younger and kinematically colder than the ``Thick Disc'' populations. This is shown in Fig. \ref{kin}, which shows different projections of our sample in velocity-age space, as well as orbital parameters as a function of age (more discussion in Chiappini et al., in prep.). 

{\it The Outer Disc population:} The green squares and circles in Fig. \ref{harps0} correspond to the metal-poor thin disc ([Fe/H] $\sim-0.5$). Apart from metallicity, its main abundance differences with respect to the bulk of the chemical thin-disc population are: 1. a light elevation in all [$\alpha$/Fe] ratios, as a consequence of the later onset of star formation in the outer disc, where this population is most likely to originate from (e.g. \citealt{Nordstroem2004, Anders2014, Hayden2015}; see also kinematic diagnostics in Fig. \ref{kin}), 2. a slight underabundance of [Sr/Fe] and [Y/Fe] with respect to the thin-disc population, but solar-like second s-process peak abundances, and 3. 
a hint for a systematic r-process ([Eu/Fe]$\sim0.2$) enhancement (consistent with that in [$\alpha$/Fe] with respect to the local disc.

{\it The young locally-born disc:} The grey circles in Fig. \ref{harps0} denote a population that we call young local disc, because 1. they are among the youngest stars ($\sim1$ Gyr; see Fig. \ref{age}), 2. they follow the local rotation curve with a very low velocity dispersion (see Fig. \ref{kin}, top panels). 
The young local disc stars have near-solar metallicities and are slightly deficient in [Mg/Fe], [Al/Fe], and [Si/Fe] with respect to the Sun and the rest of the thin disc, as expected for young stars from the stronger contribution of SN Ia yields. They are also systematically deficient in [Cu/Fe] and [Zn/Fe], while being moderately in enhanced ([s/Fe]$\sim0.15$) in s-process elements (see Figs. \ref{harps3} and \ref{age}. In fact, these stars occupy the upper boundary in the [Ba/Fe]-age and [Ba/Y]-age relations of the thin disc (see also \citealt{Spina2018}), but are still roughly consistent with 
the trends set by chemical-evolution models with metallicity-dependent yields from intermediate- and low-mass AGB stars \citep[e.g.][]{Cristallo2009, Cristallo2015, daSilva2016, DelgadoMena2017}.

{\it The remaining thin-disc component:} The black dots in Fig. \ref{harps0} stand for the remaining parts of the low-[$\alpha$/Fe] solar-vicinity disc (7 kpc $\lesssim R_{\rm guide}\lesssim9$ kpc). The morphology of this population in the t-SNE map confirms that this reference ``thin disc'' is not be a homogeneous monolithic population either (possibly more substructure could be defined, although less robustly; see Fig. \ref{mctest}). But within the scope of this paper, we define the thin disc as a broad component that has a wide range of ages and birth places, and therefore could in principle also cover a wider range of chemical abundances. Fig. \ref{harps3} shows, however, that our ``thin disc'' population, while covering a considerable metallicity range from -0.3 to 0.24, has quite small spreads in each elemental abundance relative to iron, and closely follows the solar abundance pattern. This suggests that the chemical evolution of the interstellar medium in the disc near must have been slow and very homogeneous for the past $\lesssim10$ Gyr (e.g. \citealt{Nissen2016}). The significant spread in the age-metallicity relation of the solar-neighbourhood thin disc (Fig. \ref{age}) can be explained by the presence of a strong radial metallicity gradient, together with radial mixing \citep[e.g.][Minchev et al., in prep.]{Haywood2006, Minchev2013, Anders2017a}.

\subsection{Chemically peculiar stars}\label{peculiar}

\begin{figure}\centering
 \includegraphics[width=0.5\textwidth]{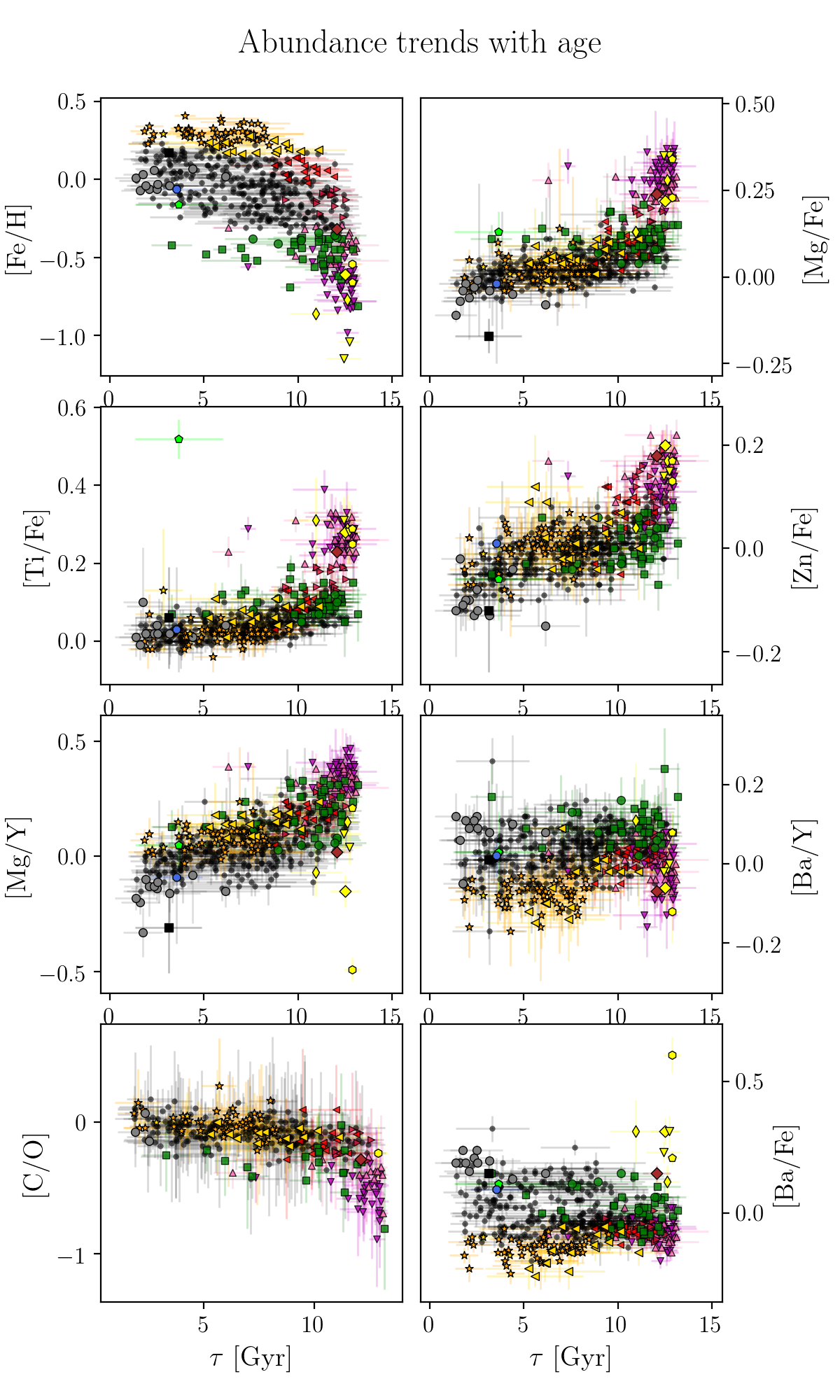}
\caption{Abundance trends of the HARPS-GTO abundances with stellar age, measured with the \texttt{StarHorse} code \citep{Queiroz2018}.}
\label{age}
\end{figure}

\begin{figure}\centering
 \includegraphics[width=0.49\textwidth]{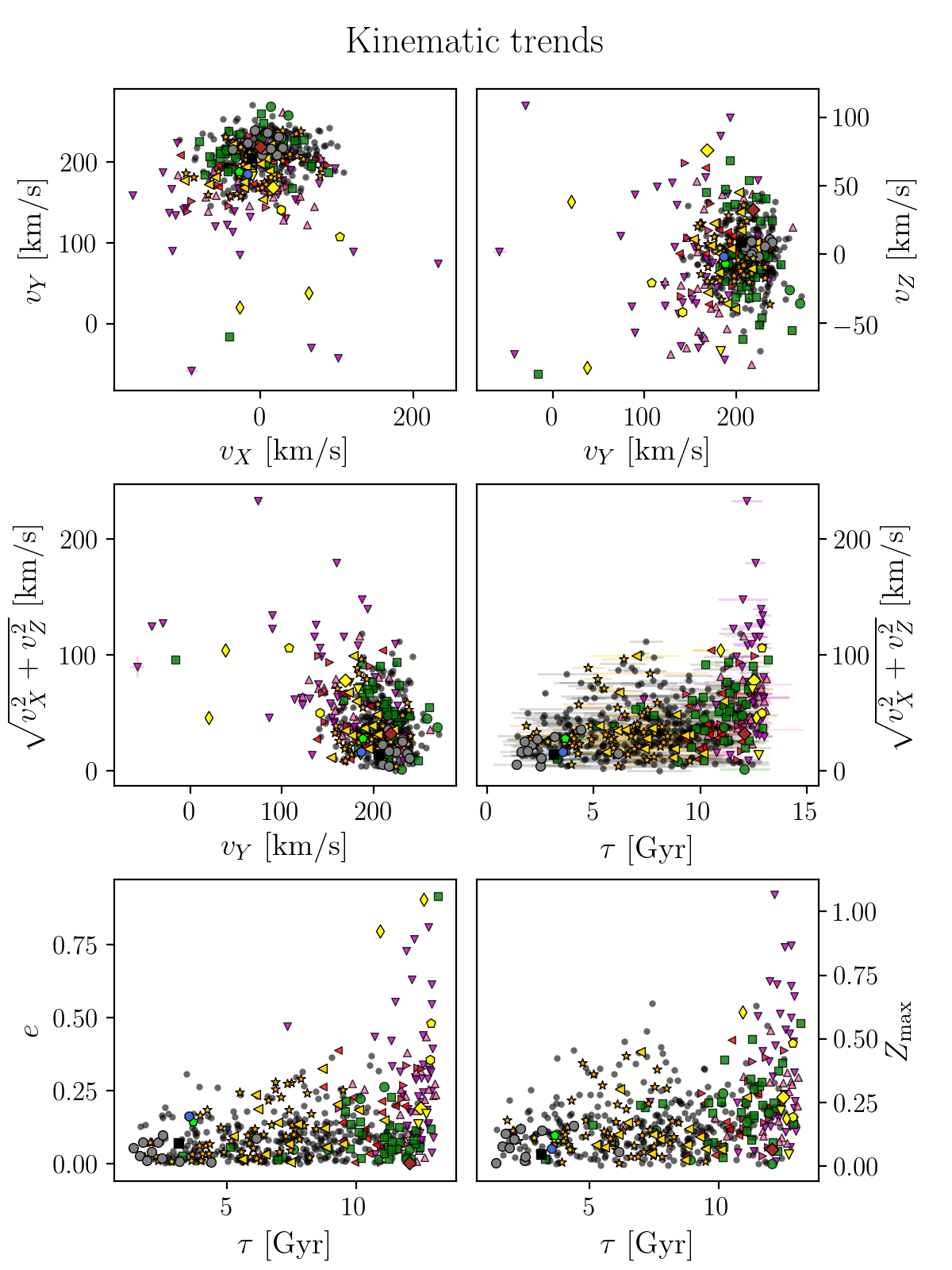}
\caption{Kinematic trends of the HARPS-GTO sample. The upper panels show the distribution of the sample in the $v_X-v_Y$ (or $UV$) and $v_Y-v_Z$ ($VW$) planes. We note that the $1\sigma$ errorbars are in most cases smaller than the symbols. The middle left panel shows the classic Toomre diagram \citep[e.g.][]{Feltzing2003}, and the middle right panel a diagram of orbital heating (as measured by $v_X$ and $v-Z$) as a function of age. The bottom panels display the orbital eccentricities and maximum heights above the Galactic plane, respectively, as a function of age.}
\label{kin}
\end{figure}

In addition to the main disc populations discussed in the previous subsection, Fig. \ref{harps0} also highlights a number of outliers and chemically peculiar stars revealed by the t-SNE projection. Some of them are known peculiar objects, some are solid, and some are dubious candidates. Their abundance patterns relative to iron are shown in the bottom panel of Fig. \ref{harps3}. Here we discuss each of them briefly.

{\it s-process-enhanced stars:} Our method clearly singles out a small group of stars with dwarf-galaxy- or globular-cluster-like, and s-process-enhanced abundance patterns (a few more were lost due to the temperature and abundance quality cuts). These seven stars (yellow points in Fig. \ref{harps0}) are all enhanced in the measured s-process elements with respect to both the thin and thick disc populations (see Fig. \ref{harps3}, bottom panel). They are also old (see Fig. \ref{age}), enhanced in [$\alpha$/Fe] -- although there is considerable star-to-star variance--, and all of them are [Al/Mg]-poorer than the thick-disc populations, placing them in an abundance regime somewhere in-between Galactic thick disc, the halo, and massive dwarf-galaxies. 

The most extreme abundance outlier in this group, as already noted by \citet{DelgadoMena2017}, is HD11397 (yellow hexagon), which shows the highest s-process abundances of the entire sample ([s/Fe]$\sim0.7$). It was classified as a so-called mild barium star by \citet{Pompeia2008} who also showed that its s-process abundance pattern is compatible with typical AGB stellar yields, possibly accreted from an unseen companion. 
Another star that was noted as a mildly s-enhanced thick-disc star by \citet{DelgadoMena2017}, is HD126803 (yellow square). The last mild s-enhancement candidate of \citet{DelgadoMena2017}, CD-436810, did not satisfy our $T_{\rm eff}$ criterion, and was therefore not included in our analysis. 

{\it s-process-enhanced stars with halo kinematics:} HD175179 (yellow pentagon) as well as BD+083095 and CD-4512460 (yellow diamonds) are mildly [s/Fe]-enhanced old halo-kinematic stars with very similar abundance patterns and metallicities between -0.66 and -0.86. BD+083095 and CD-4512460 are shown with the same symbol in all figures because they also have similar kinematics. 

{\it A high-confidence s-process-enhanced abundance pair:} The yellow triangles in all figures correspond to the nearby high proper-motion stars HD91345 and HD126681.\footnote{Both stars were observed by various solar-vicinity spectroscopic surveys such as RAVE \citep{Steinmetz2006, Kunder2017} and GCS \citep{Nordstroem2004, Casagrande2011}, resulting in compatible spectroscopic parameter determinations (although of lower quality). Using high-resolution spectroscopy, \citet{Bensby2014} measured a slightly lower metallicity for HD126681 ([Fe/H]$=-1.3$), but a very similar abundance profile to the HARPS one analysed here.} We find that the HARPS-derived abundances of \citet{DelgadoMena2017} are so similar for these two stars that they can be considered abundance-ratio twins (see Table 1). With the exception of metallicity ($2\sigma$-deviation), all [X/Fe] abundances are consistent with each other within the respective $1\sigma$ uncertainties. They have the two highest [Si/Fe] enrichments of the sample. In connection with their similar ages and space velocities (except for the discrepant $v_Z$ component and the slightly different [Fe/H] abundances), and considering the rareness of [s/Fe]-enhanced metal-poor disc stars, we propose that the two stars could have been born in the same stellar system (possibly a massive globular cluster or a dwarf galaxy) that has long since been disrupted by the Milky Way (see e.g. \citealt{Bekki2016}).

\begin{table}
\label{twintable}
\centering
\caption{Details of the s-process-enhanced abundance-ratio pair HD91345 and HD126681.}
\begin{tabular}{lcc}
Property & HD91345 & HD126681 \\
\hline \hline
S/N$_{\rm HARPS}$ & 160 & 244 \\
$T_{\rm eff}$ & $5658\pm39$ K & $5570\pm34$ K \\
$\log g_{\rm HIP}$ & $4.38\pm0.08$ & $4.64\pm0.08^{\dagger}$ \\
{[Fe/H]  }& $-1.04\pm0.03$ & $-1.15\pm0.03$ \\
{[Mg/Fe] }& $0.35\pm0.05$ & $0.35\pm0.04$ \\
{[Al/Fe] }& $0.16\pm0.01$ & $0.17\pm0.01$ \\
{[Si/Fe] }& $0.37\pm0.04$ & $0.39\pm0.05$ \\
{[Ca/Fe] }& $0.32\pm0.09$ & $0.36\pm0.06$ \\
{[TiI/Fe]} & $0.28\pm0.08$ & $0.31\pm0.06$ \\
{[Cu/Fe] }& $0.29\pm0.05$ & $0.33\pm0.03$ \\
{[Zn/Fe] }& $0.15\pm0.07$ & $0.14\pm0.07$ \\
{[Sr/Fe] }& $0.25\pm0.07$ & $0.24\pm0.05$ \\
{[Y/Fe]  }& $0.31\pm0.05$ & $0.25\pm0.07$ \\
{[ZrII/Fe]} & $0.55\pm0.06$ & $0.53\pm0.05$ \\
{[Ba/Fe]} & $0.31\pm0.06$ & $0.23\pm0.04$ \\
{[Ce/Fe]} & $0.32\pm0.08$ & $0.27\pm0.03$ \\
{[Nd/Fe]} & $0.27\pm0.06$ & $0.28\pm0.09$ \\
{[Eu/Fe]} & $0.44\pm0.32$ &  \\
\hline
Mass & $0.762^{+0.004}_{-0.011}\ M_{\odot}$  & $0.711^{+0.008}_{-0.006}\ M_{\odot}$ \\
Age & $12.9^{+0.2}_{-0.8}$ Gyr & $12.7^{+0.4}_{-1.0}$ Gyr \\
Distance & $59\pm1$ pc & $54.3\pm0.6$ pc\\
$v_X$ & $11.5\pm0.7$ km/s & $10.3\pm0.1$ km/s \\
$v_Y$ & $185.2\pm0.3$ km/s & $182.5\pm0.7$ km/s \\
$v_Z$ & $-7.5\pm0.1$ km/s & $-70.2\pm0.5$ km/s \\
\hline\hline
\end{tabular}
\tablefoot{$^{\dagger}$The high $\log g$ value results from a slightly overestimated Hipparcos parallax ($21\pm 1$ mas) compared to Gaia DR1 ($16.8\pm 0.9$ mas) and DR2 ($17.78\pm 0.07$ mas; \citealt{GaiaCollaboration2018}). The {\tt StarHorse} age for this star is driven by the higher precision of the {\it Gaia} measurements, and the posterior $\log g$ is more in line with stellar-evolution expectations for an old main-sequence star.}
\end{table}

{\it Another debris candidate at higher metallicity:} HD28701 (brown diamond in Fig. \ref{harps0}) is another interesting object with similar s-process enhancements as the [s/Fe]-enhanced stars discussed above, but at higher metallicity ([Fe/H]$=-0.32$). Like the group of yellow stars, it shows mildly enhanced ([s/Fe]$\sim0.2$) abundances of Sr, Y and Zr when compared to thick-disc stars of similar metallicity, and not as much enhancement in the second s-process peak elements Ba, Ce and Nd. It is also enhanced in the r-process element europium ([Eu/Fe]$=0.30\pm0.06$). \citet{Bensby2014} and \citet{Battistini2016} report very similar abundances for this star.

{\it High-[Ti/Fe] candidate:} HD14452 (limegreen pentagon, S/N$_{\rm HARPS}=89$) has possibly the most extraordinary abundance pattern of the \citet{DelgadoMena2017} sample: It has a metallicity of [Fe/H]$=-0.16\pm0.02$ and seems to be highly enriched in the heavier $\alpha$-elements titanium and calcium ([TiI/Fe]$=0.52\pm0.05$, [Ca/Fe]$=0.37\pm0.10$), while being only slightly enhanced in [Mg/Fe], and not at all in [Si/Fe]. Also the elevated [Al/Mg] ratio is puzzling. 
However, a reanalysis of the equivalent widths used for the TiI abundance determination has shown that the high [TiI/Fe] value measured in \citet{DelgadoMena2017} should be revised to a lower value (Delgado-Mena, priv. comm.). If an abundance pattern with extreme Ca and Ti enrichment like the one for HD14452 shown in Fig. \ref{harps3} were confirmed, this would have made the star a very interesting object: this abundance pattern would suggest a peculiar chemical enrichment, e.g. by a $\sim15\ M_{\odot}$ type-II supernova that did not produce light $\alpha$ elements but large amounts of Ca and Ti (e.g. \citealt{Ritter2018}, Fig. 26). A further interesting point is its rather young age ($\sim 4\pm 2$ Gyr), which would have made the star an extreme outlier to the age-[Ti/Fe] relation (Fig. \ref{age}, third panel) and a (metal-rich) candidate young [$\alpha$/Fe]-rich star \citep{Chiappini2015a, Martig2015}. A plausible explanation for such an object could be that it is the surviving secondary star of an old binary system whose much more massive primary exploded in a type-II supernova that polluted the atmosphere of the companion. 


{\it Low-[Mg/Fe] candidate:} HD113513 (black square) is not really an outlier in the t-SNE map, but the star with the lowest [$\alpha$/Fe] ratios of the sample, and it also sticks out in several of the abundance diagrams shown in Fig. \ref{harps0}. Its abundance profile is similar to the ``young local disc'' population defined above, except for its higher metallicity and the elevated [Al/Mg] ratio. We note the low signal-to-noise ratio (S/N$_{\rm HARPS}=26$) of the spectrum, and the consequently higher abundance uncertainties, which make this star a lower-confidence outlier.

{\it High-[Al/Mg] candidate:} HD29428 (blue octogon) is most likely not a true oddball, but a typical youngish thin-disc star with a very uncertain Al measurement ([Al/Fe]$=0.43\pm0.46$) that ended up as an outlier in the t-SNE map because the method cannot account for heteroscedastic errors.

\section{Discussion and conclusions}\label{conclusions}

The solar vicinity comprises a well-established mixture of stellar populations, among them halo stars, thick- and thin-disc stars, stars in streams, stars passing by on eccentric orbits, stars on circular orbits that have radially migrated, chemically peculiar stars, and even stars with possibly extragalactic origins (e.g. members of disrupted dwarf galaxies or globular clusters). 
In this paper we have demonstrated the use of the dimensionality reduction algorithm t-SNE to better define subpopulations in abundance space. While the non-parametric non-linear behaviour of the technique makes it difficult to estimate the significance of found subgroups or clusters, we have verified that our results depend little on the t-SNE parameter choices and are robust to abundance errors. As in other differential abundance studies, it is important to confine the analysis to narrow regions in atmospheric-parameter space to avoid spurious abundance trends induced by differences in atmospheric parameters. The t-SNE method could in principle even be coupled to a genuine cluster finding algorithm. 

Our approach allowed us to define chemical subpopulations in the solar vicinity in a more reliable way than by just looking at 2D abundance diagrams. The gap between the chemical thin and thick discs is much more prominent, as is the separation between the genuine thick-disc and the high-$\alpha$ metal-rich population. The high-[$\alpha$/Fe] population may even be composed of more than two distinct populations, but this affirmation is not as robust to abundance uncertainties. The metal-rich end of the high-$\alpha$ metal-rich population and the super-metal-rich thin-disc stars are not clearly separated in our t-SNE map, which suggests that their chemical evolution is connected (both have origins in the inner disc, \citealt[][see also Chiappini et al., in prep.]{Adibekyan2012, Haywood2018}). 

We also re-characterise the chemical thin-disc component, even excluding the metal-poor and and super-metal-rich parts corresponding to stars originating from the outer and inner disc, respectively. This broad component still covers a considerable metallicity range ($\sim-0.3$ to $+0.25$) and a wide range of ages ($\sim10$ Gyr), but has surprisingly similar abundances as the Sun, and very small spreads in the $\alpha$ abundances ($\sigma[\alpha_i$/Fe]$\simeq0.03$ -- smaller than the individual abundance uncertainties except for [Mg/Fe]), while [s/Fe] abundances dispersions ($\sim0.08$ dex) are slightly higher than the observational uncertainties (e.g. $\sim 0.04$ dex for [Ba/Fe]) because s-process elements are more sensitive to age and birth radius. In accordance with previous literature, we attribute these facts to a slow and homogeneous chemical evolution in the disc that is mainly characterised by a negative radial metallicity gradient, as well as strong radial migration that brought stars from various Galactic radii into the solar vicinity.

We found several chemically peculiar stars and candidates, of which most are s-process enhanced stars. Other outliers, such as the intriguing high-[Ti/Fe] candidate HD14452, are more likely to be due to erroneous abundance measurements. 

Our identification of the s-process-rich abundance pair HD91345/HD126681 in Sec. \ref{peculiar} demonstrates the potential of abundance-space t-SNE for chemical tagging. The viability of t-SNE for strong chemical tagging (finding dispersed members of open clusters) is still not completely clear, though. The GALAH results of \citet{Kos2018} suggest that it is possible to recover a large fraction of open clusters with abundance-space t-SNE, and to even find extratidal cluster members with this technique. On the other hand, the recent APOGEE paper by \citet{Ness2018} adopts a more pessimistic view on strong chemical tagging: the authors find that most 0.03-dex level abundance pairs (at solar metallicity) were probably not born in the same cluster, but are rather ``doppelg\"anger-abundance'' stars than actual twins. Our abundance pair is arguably a rarer case than a solar-abundance pair, and may well be a real abundance-ratio twin, as also suggested by the very similar (and precise) ages, as well as  $v_X$ and $v_Y$ space velocities. The slightly (2$\sigma$) discrepant [Fe/H] abundances could be explained naturally if the progenitor system was a massive ($\omega$ Cen-like) globular cluster or a dwarf galaxy. The question why the $v_Z$ velocities are so different remains to be resolved, as well as the question if we can find more stars with similar abundance patterns.

\bibliographystyle{aa}
\bibliography{FA_library}

\begin{acknowledgements}
We thank the referee for comments and suggestions that helped to improve the quality of this work.
FA would like to thank Elisa Delgado-Mena for sharing the re-reduced HARPS-GTO data prior to publication and for carefully reinspecting the spectrum of HD14452. He also thanks Katia Cunha, Ivan Minchev, Paula Jofr\'e, Bertrand Lemasle and the other participants of the IAU symposium 334 in Potsdam, as well as David W. Hogg and Roland Drimmel, for encouragement and critical thoughts. 

\end{acknowledgements}

\end{document}